\documentclass[aps,prl,twocolumn,amssymb,superscriptaddress]{revtex4}
\pdfoutput=1
\usepackage{subfigure}
\usepackage{graphicx}
\usepackage{rotating} 
\usepackage{color}
\usepackage{lscape}
\usepackage{rotating}
\usepackage{amsmath,amsthm,placeins}
\usepackage{epstopdf}
\usepackage[version=3]{mhchem} % Formula subscripts using \ce{}
\begin{document}

\title{Heterogeneous morphology and dynamics of polyelectrolyte brush condensates in trivalent counterion solution}

%\author{%
%Lei Liu and Changbong Hyeon}%
%\thanks{To whom correspondence should be addressed. Email: hyeoncb@kias.re.kr}
%\affiliation{%
%School of Computational Sciences, Korea Institute for Advanced Study, 85 Heogiro Dongdaemun-Gu, Seoul 02455, Republic of Korea} 

\author{Lei Liu}
\affiliation{%
Korea Institute for Advanced Study, 85 Heogiro Dongdaemun-Gu, Seoul 02455, Republic of Korea} 
\author{Philip A. Pincus}
\affiliation{%
Materials and Physics Departments, University of California at Santa Barbara, Santa Barbara, California 93106, USA}
\author{Changbong Hyeon}
\thanks{To whom correspondence should be addressed. Email: hyeoncb@kias.re.kr}
\affiliation{%
Korea Institute for Advanced Study, 85 Heogiro Dongdaemun-Gu, Seoul 02455, Republic of Korea} 

% Affiliation must include:
% Department name, institution name, full road and district address,
% state, Zip or postal code, country

%\history{%
%}
\begin{abstract}
Recent experiments have shown that trivalent ion, spermidine$^{3+}$, can provoke lateral microphase segregation in DNA brushes. 
Using molecular simulations and simple theoretical arguments, 
we explore the effects of trivalent counterions on polyelectrolyte brushes.
At a proper range of grafting density, polymer size, and ion concentration, 
the brush polymers collapse heterogeneously into octopus-like surface micelles. 
Remarkably, the heterogeneity in brush morphology is maximized and the relaxation dynamics of chain and condensed ion are the slowest at the 1:3 stoichiometric concentration of trivalent ions to polyelectrolyte charge. 
A further increase of trivalent ion concentration conducive to 
a charge inversion elicits modest reswelling and homogenizes the morphology of brush condensate.  
Our study provides a new insight into the origin of the diversity in DNA organization in cell nuclei as well as the ion-dependent morphological variation in polyelectrolyte brush layer of biological membranes.
\end{abstract}

\maketitle

%\section*{INTRODUCTION}
Polyelectrolyte brushes are ubiquitous in biological systems. 
As the main component of outer membrane in Gram-negative bacteria, 
lipopolysaccharides, consisting of  
charged O-polysaccharides side chains, form a brush layer and mediate the interaction of bacteria with their environment \cite{2009TanakaS671}.  
For vertebrates, 
negatively charged polysaccharide hyaluronic acids 
play a vital role in the organization of pericellular matrix \cite{2012Richter1466,2004Addadi1393}. 

Double-stranded DNA brushes  
on a biochip \cite{2014Karzbrun829}
at a cell-like density ($\sim 10^{4}$ kb/$\mu\text{m}^{3}$) 
have been developed as a platform to study cell-free gene expression \cite{2014BarZiv1912}. 
Due to the negative charges along the backbone, 
this synthetic system behaves like a well-defined strong polyelectrolyte brush \cite{2014BarZiv1912}, the height of which can be varied by modulating the ionic strength of monovalent salt (NaCl) solution and the grafting density \cite{2013BarZiv4534}. 
Recently, it has been reported that trivalent counterions, spermidine$^{3+}$ (\ce{Spd^3+}), can induce a collapse transition of DNA brush 
into fractal-like dendritic macroscopic domains \cite{2014BarZiv4945}.
The heterogeneous morphology of DNA condensate 
points to polyamine-mediated regulation of local DNA configuration and gene expression \cite{2011Cherstvy9942,2012Barbi9285}, underscoring the importance of understanding the effects of multivalent counterions on polyelectrolyte brush \cite{2016Tirrell284,2016Borisov402}. 
Compared with uncharged polymer brushes in good or poor solvent, 
the electrostatic interaction and osmotic pressure, 
which are readily controlled by the salt concentration and pH, 
yield additional flexibilities in manipulating polyelectrolyte brushes, 
and thus holding promise for its wide applications \cite{2014Borkovec2479,2016BarZiv142}. 
 
Conventional theoretical approaches, successful in explaining the behaviors of polyelectrolytes in monovalent salt solutions \cite{1991Pincus2912,1994Zhulina3249,2016Borisov402}, are of limited use, particularly, when multivalent counterions are at work. 
For example, the solution to the mean-field Poisson-Boltzmann theory \cite{netz2000EPJE} fails to account for phenomena such as ion condensation, charge over-compensation, and collapse of like-charged polymers 
\cite{2000May199,2008Hsiao7347,2002Shklovskii329,2003Wong8634}. 
A prediction from a nonlocal density-functional theory that 
the exponent $\nu$ in the power-law relation between brush height ($H$) and counterion concentration ($H\sim(C_{n+})^{-\nu}$) decreases with the valence of salt ion ($n$) \cite{2008Wu7713} contradicts to the finding from a recent MD simulation \cite{2014Hsiao2900} that the transition is sharper for counterions with larger valence, i.e., $\nu$ increases with $n$. 

Polyelectrolyte brushes, which retain charged groups along the chain backbone and release ions to the solution, have been modeled explicitly 
with fully accounted electrostatic interactions \cite{2000Seidel2728}. 
The dependence of brush height on counterion concentration and grafting density has been systematically investigated  
in mono-, di-, and trivalent ion solutions \cite{2014Hsiao2900}. 
Atomistic simulations of DNA arrays solvated with \ce{Na^+}, \ce{Mg^2+}, and spermine$^{4+}$ 
suggested that the transient ``bridging'' of multivalent cations 
drives the collapse of DNA \cite{2016Aksimentievgkw081}.
It was also shown that an excess trivalent counterions over-compensating the polyelectrolyte charges elicits re-entrance transition (decondensation) of brush  \cite{2006Hsiao148301,2015Chaikin61}. 

The presence of lateral instability (or morphological heterogeneity) in polyelectrolyte brush, once noticed with AFM images \cite{2010Won2021}, is evident in DNA (strong polyelectrolyte) brush condensates in polyamine solution \cite{2014BarZiv4945}. 
Simulations and theories have predicted the possibilities of lateral microphase segregation 
in homopolymer brush in \emph{poor} solvent condition for decades \cite{Lai1992JCP,grest1993Macromolecules,1993Williams1313,zhulina1998JCP,2009Dobrynin13158,2014Sommer104911,2005Pereira214908,2010Szleifer5300}, 
but the effects of multivalent ions on the morphological heterogeneity of polyelectrolyte brushes have not been fully investigated. 
Here, we carried out molecular simulations of strong polyelectrolyte brushes in trivalent counterions 
to study (1) the conditions for the heterogeneous collapse of brush and  
(2) the dynamic characteristics of polymers and multivalent ions in brush condensates.
\\

\section*{MODEL AND METHODS}
\begin{figure}[th]
\centering\includegraphics[width=1.\columnwidth]{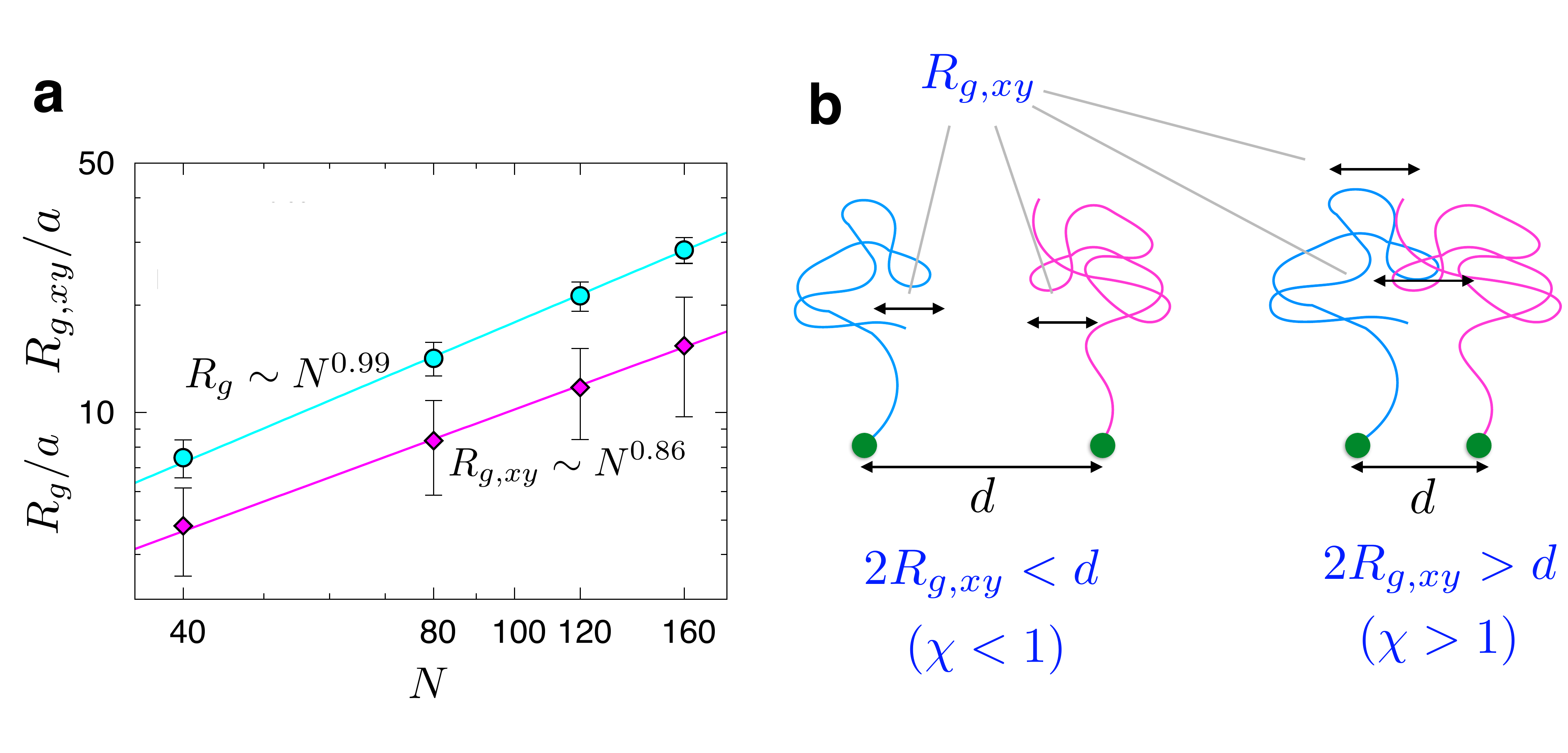}
\caption{
(a) Radius of gyration $R_{g}$ and its in-plane component $R_{g,xy}$ obtained from simulations of a single grafted polyelectrolyte chain with varying lengths $N=40$, 80, 120, and 160, grafted on surfaces.   
$R_{g}^{2} = \frac{1}{2 (N+1)^{2}} \sum_{i=0}^{N} \sum_{j=0}^{N} \sum_{\alpha=1}^{3} ({r}_{i \alpha}-{r}_{j \alpha})^{2}$ and 
$R_{g,xy}^{2} = \frac{1}{2 (N+1)^{2}} \sum_{i=0}^{N} \sum_{j=0}^{N} \sum_{\alpha=1}^{2} ({r}_{i \alpha}-{r}_{j \alpha})^{2}$,   
where $r_{i \alpha}$ is the $\alpha$-th Cartesian component of monomer $i$. 
The scaling relationship of $R_{g,xy}^2\sim N^{1.7}$ acquired from simulations is used to discuss the scaling arguments made in Eqs.\ref{eqn:scaling1}, \ref{eqn:scaling2}.
(b) The chain overlap parameter $\chi$ is illustrated. 
}
\label{brushmodel}
\end{figure}

{\bf Model.} 
A well tested coarse-grained model of strong polyelectrolyte brush 
\cite{2000Seidel2728,2003Stevens3855,2014Hsiao2900} was used in our study. 
Total $M$ polymer chains 
were grafted to the uncharged surface at $z=0$, 
in a 2D triangular lattice with a lattice spacing $d$ (Fig.~\ref{brushmodel}a). %\ref{contrast} 
Each chain, consisting of $N$-negatively charged monomers and a neutral monomer grafted to the surface, was initially neutralized using $N$-monovalent counterions. 
The simulation box has a dimension of 
$L_{x}\times L_{y}\times L_{z} = {(\sqrt{M} d)} \times {(\sqrt{3M} d/2)} \times {(2 N a)}$, 
where $a$ is the size (diameter) of chain monomer. %ion particles and chain monomers.   
Periodic boundary conditions were applied in $x$ and $y$ dimensions, 
and an impenetrable neutral wall was placed at $z=2 N a$. 
The location of the wall is far from the brush region, so that 
the density profiles of ions in the bulk region are effectively constant 
\cite{2004Seidel16870,2005Seidel9341}.

%Fig.1
\begin{figure}[th]
\centering\includegraphics[width=1.\columnwidth]{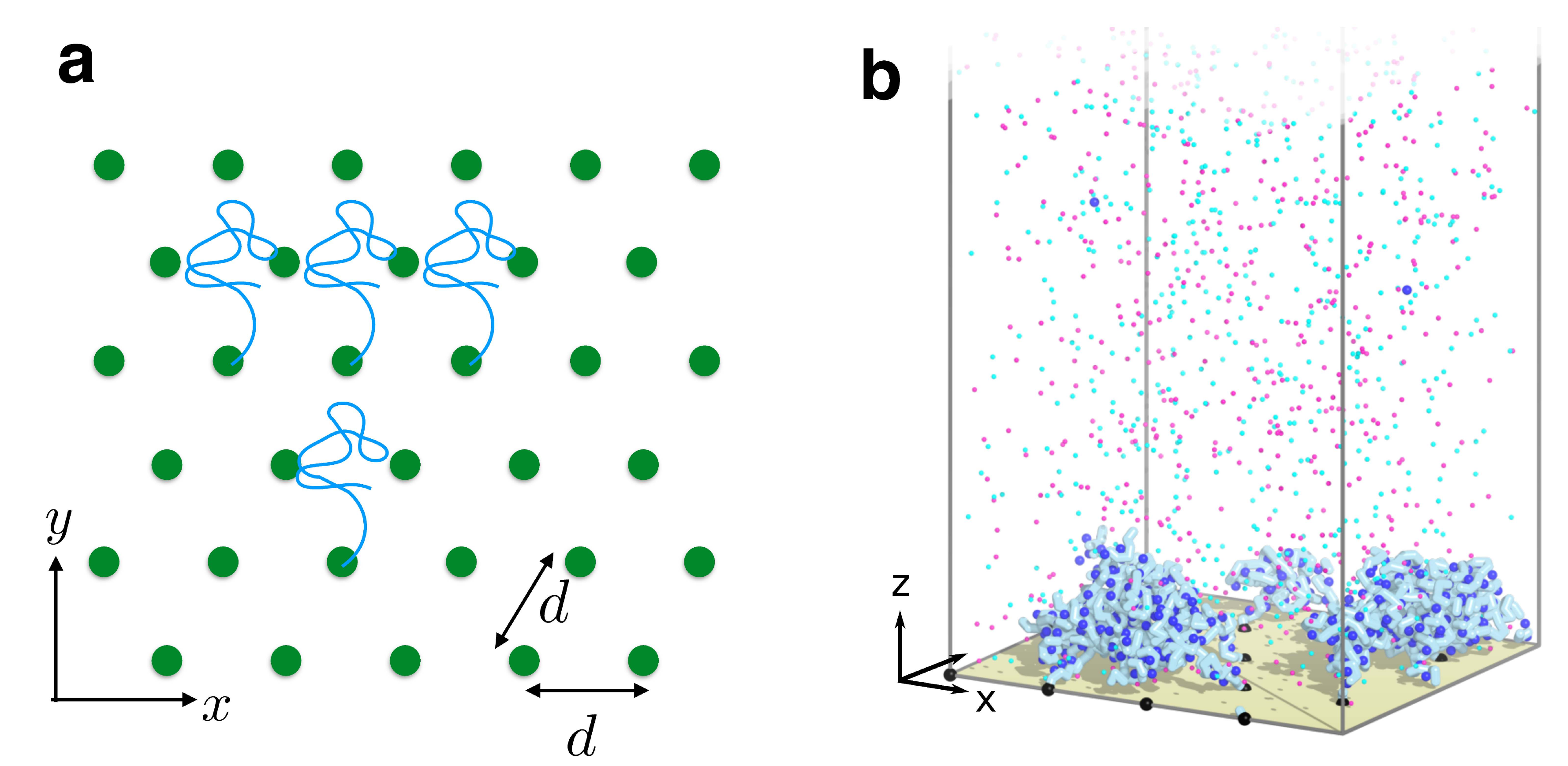}
\caption{
(a) Geometry of brush simulation. Each brush polymer is grafted in a 2D triangular lattice with spacing $d$. 
(b) A snapshot from simulation. Shown is the bottom part of simulation box where $4 \times 4$ polyelectrolyte brush (light blue) is collapsed by trivalent cations (deep blue). The monovalent counterions (magenta) are released from the polymer brush into the bulk. 
The monovalent anions are shown in cyan. 
}
\label{model}
\end{figure}

In order to model a flexible strong polyelectrolyte brush in good solvents with multivalent salts, we considered the following energy potentials. 
First, the finite extensible nonlinear elastic bond potential 
was used to constrain the bond distance between the neighboring chain monomers, 
\begin{equation}
U_{bond}(r) = -\frac{k_{0} R_{0}^{2}}{2} \log\left(1-\frac{r^{2}}{R_{0}^{2}}\right),
\end{equation}
with a spring constant $k_{0}$ and a maximum extensible bond length $R_{0}$. 
Second, the excluded volume interaction was modeled in between ions and chain monomers 
by using the Weeks-Chandler-Andersen potential 
\begin{equation}
U_{excl}(r) = 4 \epsilon \left[\left(\frac{a}{r}\right)^{12}-\left(\frac{a}{r}\right)^{6}+\frac{1}{4}\right],
\end{equation}
when $r \leq 2^{1/6} a$; otherwise, $U_{excl}(r)=0$. 
Third, we assigned the Columbic interactions between charged particles $i$, $j$  
\begin{equation}
U_{elec}(r) = \frac{k_{B} T \lambda_{B} q_{i} q_{j}}{r},
\end{equation} 
where $q_{i,j}$ is the corresponding charge in the unit of elementary charge $e$, 
and $\lambda_{B}=e^{2}/(4 \pi \epsilon_{0} \epsilon_{r} k_{B}T)$ 
is the Bjerrum length with the vacuum permittivity $\epsilon_{0}$.  
We used $\epsilon_{r}\approx 80$ for the relative dielectric constant of the solvent. 
Except for the anchored chain monomers, we considered a repulsion between the wall and any particle whose distance from the wall satisfies $z\leq 0.5 a$, such that 
\begin{equation}
U_{wall}(z) = 4 \epsilon \left[\left(\frac{a}{z+\Delta}\right)^{12}-\left(\frac{a}{z+\Delta}\right)^{6}+\frac{1}{4}\right], 
\end{equation} 
with $\Delta = (2^{1/6}-0.5) a$; otherwise, $U_{wall}(z)=0$. 
Note that $U_{wall}(z)$ and its derivative are continuous at $z=0.5 a$. 
For simplicity, we assume that all the ions and chain monomers 
have the same values of $a$ and $\epsilon=k_{B} T$, 
which are used as the basic length and energy units in this model. 
With $k_{0}=5.83$ $\epsilon/{a}^{2}$, $R_{0}=2 a$, and $\lambda_{B}=4 a$, 
MD simulations of a single grafted polyelectrolyte chain give 
an average bond length ${\langle b \rangle} = 1.1 a$ $(\sim a)$. 
For double-stranded DNA this bond length maps to the effective charge separation ($\sim 1.7$ {\AA}) along the chain.  
Since $\lambda_{B} = 7.1$ {\AA} in water at room temperature, 
one gets $\lambda_{B} = 7.1/1.7\times a$ $(\approx 4 a)$. 
\\

{\bf Simulation Methods.} 
Simulations were carried out 
using ESPResSo 3.3.1 package \cite{2006Holm704,2013Holm1}. 
Depending on the property of interest (static or dynamic) we used two different algorithms for simulations \cite{VeitshansFoldDes97,Hyeon05PNAS,Hyeon06BJ,Hyeon08JACS}. 
In order to get a well sampled ensemble of configurations and obtain static (equilibrium) property of brush system for a given condition, 
we set the friction coefficient to a low value and integrated the following underdamped Langevin equation of motion using the velocity Verlet algorithm. 
\begin{align}
m\frac{d^2\vec{r}_i}{dt^2}=-\zeta_{\text{MD}}\frac{d\vec{r}_i}{dt}-\vec{\nabla}_{\vec{r}_i}U(\vec{r}_1,\vec{r}_2,\ldots)+\vec{\xi}(t)
\end{align}
where the random force noise, satisfying $\langle\vec{\xi}(t)\rangle=0$ and $\langle\vec{\xi}(t)\cdot\vec{\xi}(t')\rangle=6\zeta_{\text{MD}}k_BT\delta_{ij}\delta(t-t')$, was used to couple the simulated system to Langevin thermostat.
The characteristic time scale of this dynamics
is given to be $\tau_{\text{MD}} = (ma^2/{\epsilon})^{1/2}$ \cite{1992Thirumalai695}. We chose the time step of integration 
$\delta t = 0.01 \tau_{\text{MD}}$ and a small friction coefficient $\zeta_{\text{MD}} = 0.1 m \tau_{\text{MD}}^{-1}$.
A salt-free brush was first equilibrated for $1 \times 10^{6}$ $\delta t$, corresponding to $\mathcal{T}_{\text{eq}} = 10^4$ $\tau_{\text{MD}}$, 
from an initial configuration where polymer chains were vertically stretched 
with monovalent counterions homogeneously distributed in the brush region. 
To simulate the effect of trivalent cations on the brush systems, we randomly added cations and its monovalent coions (anions) into the brush-free zone. %in the Supporting Information). 
The production runs were generated for $10^{5}\times \tau_{\text{MD}}$.  
A snapshot from simulations was saved every $25\times \tau_{\text{MD}}$ for the analysis. 

When examining dynamic behaviors of system (such as configurational relaxation of brush) and evaluating the associated time scale of dynamics, 
which we report in Fig.\ref{isf} by using Eq.\ref{eq-isf}, simulations under overdamped condition is more appropriate \cite{Hyeon08JACS,Hinczewski2010JCP}. 
In order to probe the dynamic behavior of brush we switch our simulation algorithm to Brownian dynamics (BD). 
In BD simulations the following equation of motion is integrated: 
\begin{equation}
\frac{d \vec{r}_{i}}{dt} = - \frac{D_{i0}}{k_B T}{\vec{\nabla}}_{\vec{r}_{i}} U(\vec{r}_1,...,\vec{r}_N) +\vec{R}_{i}(t),
\end{equation}
where $D_{i0}$ is the bare diffusion coefficient of the $i$th particle, 
and $\vec{R}_{i}(t)$ is the Gaussian random noise satisfying the fluctuation-dissipation theorem 
$\langle \vec{R}_{i}(t) \cdot \vec{R}_{j}(t')\rangle = 6 D_{i0} \delta_{ij} \delta(t-t')$. 
For the diffusion constant of ions and chain monomers, we used 
$D_{i0} = 1.33, ~2.03, ~0.48, ~0.24 \times 10^{3}$ $\mu m^2/s$, 
for \ce{Na^+}, \ce{Cl^-}, \ce{Spd^3+} and DNA monomer respectively 
\cite{2004David16811,1996Schultz2847,2002Torre1782}. 
For the integration time step of simulation, we chose $\delta t_{\text{BD}} = 5 \times 10^{-3} \tau_{\text{BD}}$, which is much smaller than the Brownian time $\tau_{\text{BD}}=a^{2}/D$.  
The last configuration of brush obtained from MD simulation was used as the starting configuration of BD simulation.  
The total simulation time for BD in each trajectory is $7.5 \times 10^{4} \tau_{\text{BD}}$. 

Because of the slab geometry of our system, an electrostatic layer correction to 
the Particle-Particle Particle-Mesh algorithm ($\text{P}^3$M+ELC) was used 
to calculate the long-range Coulomb interactions \cite{2002Holm2496}, 
which required an extension of the simulation box along $+z$ direction 
so that there is a vacuum slab above the top impenetrable wall. 
The height of the slab was chosen to be $N a/4$, and the error tolerance was set to $10^{-5}\varepsilon$. 
Differences between the electrostatic potentials of some exemplary configurations, 
calculated by $\text{P}^3$M+ELC and by MMM2D \cite{2002Holm327} algorithm, 
were negligible $\sim 0.01 \epsilon$, 
indicating that Coulomb interactions have been fairly accounted. %{\color{blue}(dielectric surface)} 

%Fig.2

\begin{table*}
  %\begin{center}
%  \begin{scriptsize}
  \caption{\label{sysinfo}
    Parameters for the simulated brush systems. 
    Values of polymer chain length $N$, dimensionless overlap parameter $\chi$, 
    dimensionless trivalent counterion concentration $q_{3+}$, 
    grafting density $\sigma$ in units of $10^{-3} a^{-2}$, 
    and trivalent ion concentration $C_{3+}$ in units of $10^{-3} a^{-3}$. 
    }
%\end{scriptsize}
\centering
  \scriptsize
 % \begin{tabular}{p{1.0cm}p{.4cm}lllp{1.0cm} p{.8cm}p{.4cm}lllp{1.0cm} p{.8cm}p{.4cm}llll}
     %\begin{tabular}{c| c c c c c || c| c c c c c || c| c c c c c}
     \begin{tabular}{|c| c c c c c || c| c c c c c || c| c c c c c|}
    \hline
    Group & $N$ & $\chi$ & $q_{3+}$ & $\sigma$ & $C_{3+}$  &  %
    Group & $N$ & $\chi$ & $q_{3+}$ & $\sigma$ & $C_{3+}$  &  %
    Group & $N$ & $\chi$ & $q_{3+}$ & $\sigma$ & $C_{3+}$ \\
    \hline
    A & 80  & 2.0 & 0.05 & 8.30 & 0.07  &  B & 80  & 0.5 & 2.0  & 2.08 & 0.69  &  C & 120 & 1.0 & 1.0  & 2.09 & 0.35\\
    A &  80 &  2.0   & 0.1  &8.30 & 0.14  & B   &  80   &  0.5 & 4.0  & 2.08  & 1.38  &  C  &   120  & 1.8 & 1.0 & 3.76 & 0.63\\ %\cmidrule{7-12}
    A &   80&    2.0 & 0.2  &8.30 & 0.28  &  C & 40  & 1.0 & 1.0  & 12.5 & 2.08  &  C  & 120    & 2.0 & 1.0 & 4.18 & 0.70\\
    A &   80&   2.0  & 0.4  &8.30 & 0.55  &  C &  40   & 1.8 & 1.0 & 22.4 & 3.74  &  C  & 120    & 2.2 &  1.0 & 4.59 & 0.77\\
    A &   80&   2.0  & 0.8  &8.30 & 1.11  & C  &  40   & 2.0 & 1.0 & 24.9 & 4.15  &  C  & 120    & 3.0 &  1.0    & 6.27 & 1.04\\
    A &   80&   2.0  & 1.0  &8.30 & 1.38  &  C  &  40   & 2.2 & 1.0  & 27.4 & 4.57  & C   & 120    & 5.0 &1.0      & 10.4 & 1.74\\
    A &   80&   2.0  & 2.0  & 8.30 & 2.77  &  C  &  40   & 3.0 & 1.0 & 37.4 & 6.23  &  C  &  120   & 9.0 &     1.0 & 18.8 & 3.13\\
    A &   80&   2.0  & 4.0  &8.30 & 5.54  &   C & 80  & 1.0 & 1.0  & 4.15 & 0.69  &  C  & 160 & 1.0 & 1.0  & 1.22 & 0.20\\ %\cmidrule{1-6}
    B & 80  & 0.5 & 0.05 & 2.08 & 0.02  &  C  & 80    & 1.8 & 1.0 & 7.47 & 1.25  &  C  &  160   & 1.8 &1.0      & 2.20 & 0.37\\
    B &  80& 0.5    & 0.1  & 2.08 & 0.03  & C   & 80    & 2.0 & 1.0 & 8.30 & 1.38  & C   & 160    & 2.0 &     1.0 & 2.44 & 0.41\\
    B &  80& 0.5    & 0.2  & 2.08  & 0.07  & C   & 80    & 2.2 & 1.0 & 9.13 & 1.52  & C   & 160    & 2.2 & 1.0     & 2.68 & 0.45\\
    B &  80& 0.5    & 0.4  & 2.08  & 0.14  &  C  & 80    & 3.0 & 1.0 & 12.5 & 2.08  &  C  &  160   & 3.0 &  1.0    & 3.66 & 0.61\\
    B &  80& 0.5    & 0.8  & 2.08  & 0.28  &  C  & 80    & 5.0 &  1.0 & 20.8 & 3.46  & C   & 160    & 5.0 &  1.0    & 6.10 & 1.02\\
    B &  80& 0.5    & 1.0  & 2.08  & 0.35  &  C  &  80   & 9.0 &  1.0 & 37.4 & 6.23  &  C  &  160   & 9.0 &  1.0    & 11.0 & 1.83\\
    \hline
  \end{tabular}
%  \end{center}
\end{table*} 

Brushes of varying chain length ($N$) and grafting density ($\sigma$) were studied  
at varying trivalent counterion concentration $C_{3+}$. 
The parameters explored for the brush system are listed in Table \ref{sysinfo}. %\ref{sysinfo}. 
We introduce two dimensionless parameters to describe the system. 
The extent of lateral overlap between neighboring polymer chains is formulated using the chain overlap parameter 
$\chi \equiv (2 R_{g,xy}/d)^2$, where $R_{g,xy}$, which scales as $R^2_{g,xy}\sim N^{1.7}$ (Fig.~\ref{brushmodel}), is the in-plane component of the gyration radius of a \emph{single} grafted polyelectrolyte chain.  
The parameter $\chi$ can be related to the grafting density $\sigma$ via $\chi=2\sqrt{3}\sigma R_{g,xy}^2\sim \sigma N^{1.7}$. 
Collective features of brushes are expected to become significant only for $\chi \geq 1$ (see Fig.\ref{brushmodel}).%

Next, we used another dimensionless parameter $q_{3+}$ which defines the ratio of the total positive charges on the trivalent counterions in the simulation box of volume $V$ ($Q_{3+}=3C_{3+}V$) to the total negative charges on the chain monomers ($Q_{mono}=M\times N$), i.e., $q_{3+}=Q_{3+}/Q_{mono}$. 
Thus, $q_{3+}$ becomes 1 ($q_{3+}=1$) when the concentration of trivalent counterions is set to the 1:3 stoichiometric ratio at which a trivalent ion ($+3$) neutralizes 3 monomers in the brush.   

\section*{RESULTS}
\begin{figure*}[t]
\centering\includegraphics[width=1.8\columnwidth]{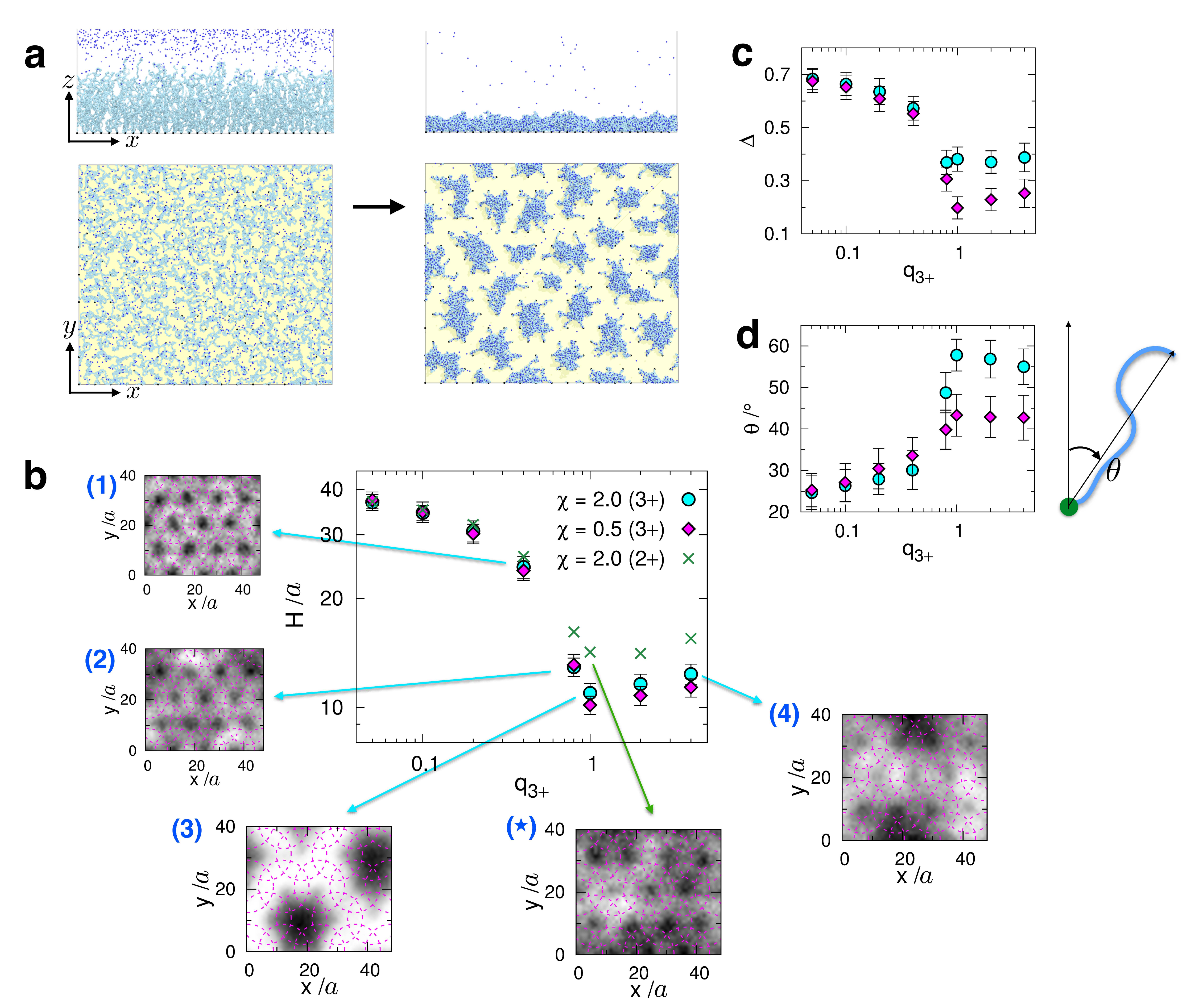}%\vspace*{-5.5pt}
\caption{
(a) Configuration of brush (side and top views) before and after trivalent counterion-induced collapse to condensates ($M=16\times16, N=80$, $\chi=2.0$ and $q_{3+}=1.0$).
(b) Brush height $H = 2~{\sum_{i=0}^{N} z_{i}}/(N+1)$. 
Lateral monomer density maps of brushes ($\chi=2.0$) averaged over time, $\langle \rho(x,y) \rangle$ are shown in the subpanels (1)--(4) at $q_{3+}=0.4$, 0.8, 1.0, 4.0. (the first monomers anchored to the surface ($z < 2 a$) are excluded from the visualization using $\langle \rho(x,y) \rangle$). 
The dashed magenta circles on the profiles depicts the radius $R_{g,xy}$ with respect to the anchoring point of chain.
The brush height with increasing divalent counterion concentration is depicted with the symbols $\times$ and monomer density map for $q_{2+}(=\frac{2C_{2+}V}{M\times N})=1.0$ is shown. 
(c) Asphericity  
$\Delta = {\sum_{i=1}^{3} (\lambda_{i} - \bar{\lambda})^2} / {6 \bar{\lambda}^2}$ where $\lambda_{i}$ are the eigenvalues of inertia tensor, 
$\bar{\lambda} = {\sum_{i=1}^{3} \lambda_{i}}/3$ \cite{1986Aronovitz1445,2006Hyeon194905}. 
$\Delta=0$ for a perfect sphere.  
(d) Polar angle $\theta = \angle(\vec{z}, \vec{r}_{e})$, where $\vec{r}_{e}$ is the end-to-end vector of individual chains,   
as a function of trivalent ion concentration $q_{3+}$. 
}
\label{betaHRAT}
\end{figure*}

{\bf Effect of trivalent ion on polymer brush.} 
Upon addition of trivalent counterions to the bulk phase, the trivalent ions are spontaneously pulled into the brush phase, replacing the monovalent counterions neutralizing the bare charges on the brush monomers (Fig.~\ref{betaHRAT}a). 
The morphology of brush condensate varies depending on the size of brush chain ($N$), grafting density ($\sigma$ or equivalently $\chi$), and trivalent ion concentration ($q_{3+}$).

\begin{figure*}[t]
\centering\includegraphics[width=1.4\columnwidth]{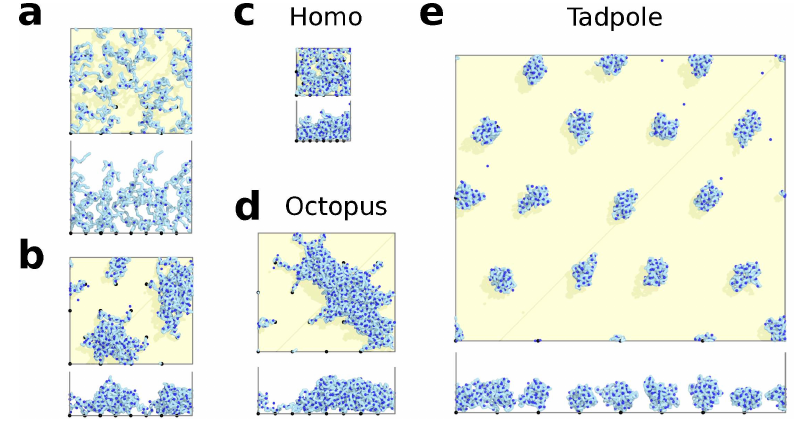}%\vspace*{-5.5pt}
\caption{
Top and side views of snapshots from simulations of polyeletrolyte brush condensates at various conditions.
(a) $N=80, \chi=2.0, q_{3+}=0.4$.
(b) $N=80, \chi=2.0, q_{3+}=1.0$.
(c) $N=40, \chi=3.0, q_{3+}=1.0$.
(d) $N=120, \chi=3.0, q_{3+}=1.0$.
(e) $N=160, \chi=1.0, q_{3+}=1.0$. 
(b), (d), and (e) correspond to the brush morphologies in (6), (4), and (3) of the structural diagram (Fig.\ref{pDiag}), respectively.
}
\label{cfgs}
\end{figure*}

For $N=80, \chi=2.0$ with an increasing trivalent ion concentration ($C_{3+}$) but with $q_{3+} < 1$ (Group A in Table \ref{sysinfo}), 
the brush height drops gradually owing to 
the decrease of the osmotic pressure \cite{2014BarZiv1912},  
and the lateral distribution of the brush becomes more inhomogeneous. 
Our study finds that at $q_{3+}=1$ the height of brush ($H$) is minimized (Fig.~\ref{betaHRAT}b) and that the grain boundary between chain-free and chain-rich phase projected on the $xy$ plane becomes clear in the same condition.
The density profiles of chain monomers in $xy$ plane $\langle \rho(x,y) \rangle$ 
(Fig.~\ref{betaHRAT}b) clearly visualize the morphology of brush condensate at various $q_{3+}$, highlighting that the development of lateral heterogeneity in the brush condensate is maximized at $q_{3+}=1$. 
At this ideal charge compensation point ($q_{3+}=1$), the six chains stretched from the corners of a hexagon are organized into aggregates entangled at the center, forming octopus-like surface micelles
(Fig.~\ref{betaHRAT}b-(3), see also a snapshot in Fig.~\ref{cfgs}b), demarcated by the grain boundary whose width is about $d = 2 R_{g,xy}/\sqrt{\chi}$. 
A further increase of $q_{3+}$ beyond $q_{3+} = 1$ ($q_{3+}>1$) elicits reswelling of the brush along the $z$-direction, obscuring the boundaries between the surface micelles (Fig.~\ref{betaHRAT}b-(4)).    
  
%Fig.3
%\begin{figure*}[h!]

%Fig.4
%\begin{figure*}[h!]

The lateral heterogeneity in brush condensate is unique to the counterions whose valence is greater than two. 
It is worth noting that as depicted in Fig.~\ref{betaHRAT}b-($\star$), addition of divalent counterions even at $q_{2+}=1$, where the divalent ions can fully neutralize the total bare charge of chain monomers, does not elicit the heterogeneous brush collapse as in Fig.~\ref{betaHRAT}b-(3).

The morphology of a sparsely grafted polyelectrolyte brush condensate ($\chi < 1$) differs greatly from that of a dense brush condensate (Group B in Table \ref{sysinfo}). 
Even at high $C_{3+}$ (or high $q_{3+}$), ion mediated inter-chain attractions are not strong enough 
to sustain a fused micelle overcoming the penalty of stretching individual polymer from the grafting point \cite{1993Williams1313}. 
As a consequence, each grafted chain collapses into a separate tadpole-like/mushroom structure 
(Figs.~\ref{cfgs}e).

Changes of the in-plane morphology of the brush are also mirrored in the configuration of individual polyelectrolyte chain. 
Of immediate notice are the non-monotonic dependences 
of brush height ($H$), chain asphericity ($\Delta$), and tilt angle ($\theta$) on $q_{3+}$ (Fig.~\ref{betaHRAT}). 
The chains in dense brush at $q_{3+} = 1$ 
display the greatest compression along the $z$-axis and tilt angle, 
which can be explained by laterally stretched configurations of the polyelectrolyte chain in the octopus-like condensate.  
Both $\Delta$ and $\theta$ are smaller 
in a sparse brush ($\chi=0.5$) than in a dense one ($\chi=2.0$), pointing to their tadpole-like configuration.

%Fig.5
\begin{figure}[h!]
\centering\includegraphics[width=.96\columnwidth]{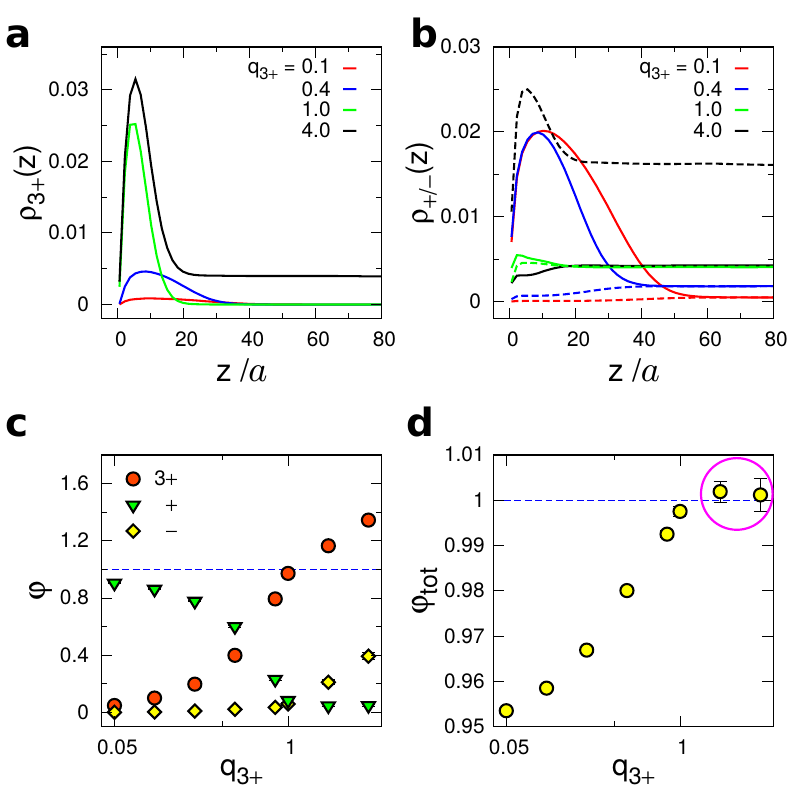}%\vspace*{-5.5pt}
\caption{
Distribution of ions.  (a) Distributions (along $z$ axis) of trivalent counterions, (b) monovalent counterions (solid lines) and coions (dashed lines), in units of $a^{-3}$,  
calculated from a collapsed brush system with $N=80$ and $\chi=2.0$ for varying $q_{3+}=0.1$, 0.4, 0.8, 1.0, and 4.0. 
(c) Ratio of total charge of trivalent counterions ($\varphi_{3+}$), monovalent counterions ($\varphi_{+}$), monovalent coions ($\varphi_{-}$), trapped within the range of Bjerrum length $\lambda_B\sim 4a$ from any chain monomer)  
to the amount of bare charges in the brush. 
(d) $\varphi_{tot}(=\varphi_{+3}+\varphi_{+1}-\varphi_{-1})$ as a function of $q_{3+}$.  
The condition of brush charge neutralization is marked with a blue dashed line,   
and the charge inversions by counterions and coions at $q_{3+}>1$ 
are highlighted with a magenta circle enclosing the two points greater than $\varphi_{tot}=1$.
}
\label{iondist}
\end{figure}

Next, of great interest are the distribution of counterions in the brush and bulk regions. 
For $q_{3+}\ll 1$, when there is only a small amount of trivalent counterions in the system, the monovalent cations are still dominant in the brush region ($\rho_{3+}(z) \ll \rho_{+}(z)$, compare the red solid lines in Fig.\ref{iondist}a and Fig.~\ref{iondist}b). 
But, as $q_{3+}(<1)$ increases, more trivalent counterions are condensed into the brush region, compressing the brush into a more compact condensate, and when the 1:3 stoichiometric condition of $q_{3+}=1$ is reached the distribution of trivalent counterions in the brush region exceeds that of monovalent counterion  ($\rho_{3+}(z)\gg \rho_{+}(z)$, compare the green lines in Fig.~\ref{iondist}a and Fig.~\ref{iondist}b). 
For $q_{3+} > 1$, 
the excess trivalent counterions are condensed 
on polyelectrolyte brush region together with monovalent anions (see the black solid/dashed lines in Fig.~\ref{iondist}a,b and symbols in Fig.\ref{iondist}c). 
When all the charge ratios of ions are summed up ($\varphi_{tot}=\varphi_{3+}+\varphi_{+1}-\varphi_{-1}$) for $q_{3+}>1$, the total charge ratio ($\varphi_{tot}$) is slightly greater than the unity, $\varphi_{tot}\gtrsim 1$ (Fig.~\ref{iondist}d). 
In fact, this charge inversion is responsible for the reswelling of brush suggested in Figs.~\ref{betaHRAT}b-d. 
Here, it is noteworthy that the density profile of trivalent ions is uniform in the polymer-free zone, which indicates that the simulation box size is sufficiently large and that the effect of box size in the $z$-direction on brush structure is negligible.  

%Fig.6
\begin{figure}[th]
\centering\includegraphics[width=.96\columnwidth]{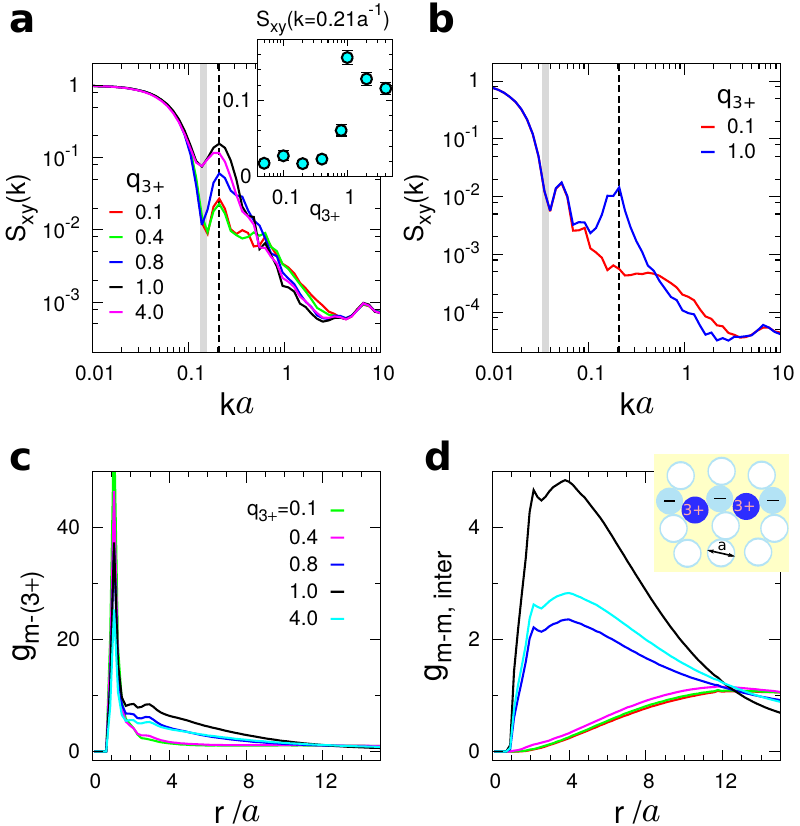}%\vspace*{-5.5pt}
\caption{
Lateral static structure factor $S_{xy}(k)$  
as a function of wave number $k$ 
calculated at $N=80$ and $\chi=2.0$ for (a) $4\times4$ and (b) $16\times16$ brush systems   
with various trivalent ion concentrations $q_{3+}=0.1,0.4,0.8,1.0,4.0$. 
Dashed vertical lines indicate the peaks around $ka=0.21$, 
whose amplitude is plotted with varying $q_{3+}$ (inset). 
The gray solid lines demarcate the range $2\pi/L_{x} \leq k \leq 2\pi/L_{y}$, i.e., the periodic boundary of the simulation box. 
Pair correlation functions between 
(c) monomer and trapped trivalent ion ($3+$) and 
(d) monomers from different chains, 
as a function of the inter-particle distance $r$. 
The insert of (d) is a cartoon of the chain ``bridging'' mechanism of trivalent ions, 
which explains the peak positions of $g_{m-m,inter}$ around $r = 2 a$ and $4 a$.
}
\label{ssf}
\end{figure}

In order to further quantify the lateral configuration of the brush, 
we calculated the in-plane static structure factor, 
\begin{equation}
S_{xy}(k) = \Big \langle \Big \langle \frac{1}{N_{m}^2} \bigg \arrowvert \sum_{i,j=1}^{N_{m}} e^{i \vec{k} \cdot (\vec{r}_i-\vec{r}_{j})} \bigg \arrowvert\Big \rangle_{|\vec{k}|} \Big \rangle,
\label{eq-ssf} 
\end{equation} 
where $N_{m} = M \times N$ is the total number of non-grafted chain monomers in the brush, 
$\vec{r}_{i}$ is the position of the $i$th monomer, and $\vec{k}$ is 2D wave vector in the $xy$ plane.
$S_{xy}(k)$ is evaluated by first integrating over the space of $| \vec{k} | = k$, 
and then averaging over the simulation trajectory.  
As shown in Fig.~\ref{ssf}a, %in the case of $q = 1.0$ (dashed line), 
$S_{xy}(k)$, calculated for $N=80$ and $\chi=2.0$, in the range of $k>2\pi/L_y$ is maximized at $ka \approx 0.21$. This indicates that at a characteristic length scale of $R\sim 2\pi/k \approx 30a$ there is a repeated pattern in the system, which corresponds to the octopus-like surface micelle domains of size $\sim 30 a$ (see the monomer density map of $\langle\rho(x,y)\rangle$ in Fig.\ref{betaHRAT}b-(3)). 

In accord with our observation that the lateral heterogeneity becomes most evident at $q_{3+}=1$ in Fig.\ref{betaHRAT}, non-monotonic dependence of $S_{xy}(k=0.21a^{-1})$ on $q_{3+}$, is again maximized sharply at $q_{3+}=1.0$ (Fig.\ref{ssf}a inset).  
At $q_{3+}=0.1$, which supposedly corresponds to a homogenous brush phase, a peak is still identified in $S_{xy}(k)$, but this is due to an artifact of periodic boundaries imposed on the $4\times 4$ brush system (Fig.~\ref{ssf}a). 
In $S_{xy}(k)$ calculated from $16\times 16$ brush system (Fig.~\ref{ssf}b),   
this peak at $q_{3+}=0.1$ vanishes, while the peak of $S_{xy}(k)$ at $ka=0.21$ in the salted brush ($q_{3+}=1.0$) still remains. 
It is worth noting that the domain size of surface micelle ($R=2\pi/k \approx 30 a$, the value of $k$ which maximizes $S_{xy}(k)$) 
is identical for $4 \times 4$ and $16 \times 16$ brush systems. %

All these non-monotonic dependences on $q_{3+}$
arise from the effective inter-chain monomer attractions mediated by the condensed trivalent counterions.
The pair correlations between the charged monomers and trivalent ions  
are strongest at $q_{3+}=1.0$ (Fig.~\ref{ssf}c). 
Furthermore, the inter-chain monomer correlation has the greatest values at $r=2a$ and $r=4a$ (Fig.~\ref{ssf}d). 
It is worth noting that the pair correlation function between trivalent ions condensed to the brush is not as sharp or structured as 
that of Wigner crystal which was observed in flexible polyelectrolyte chain in free space at relatively good solvent condition collapsed by multivalent ions \cite{NamkyungMacro01}. 
In fact, the diffusivity of the trivalent ions is finite even when they are trapped in the polyelectrolyte brush (see below).   
Still, the pronounced correlation at $r=2a$ lends support to the ``bridging'' mechanism by which charged monomers in polyelectrolyte are in effective attraction mediated by trivalent ions.    
\\

%Fig.7
\begin{figure*}[t]
\centering\includegraphics[width=1.4\columnwidth]{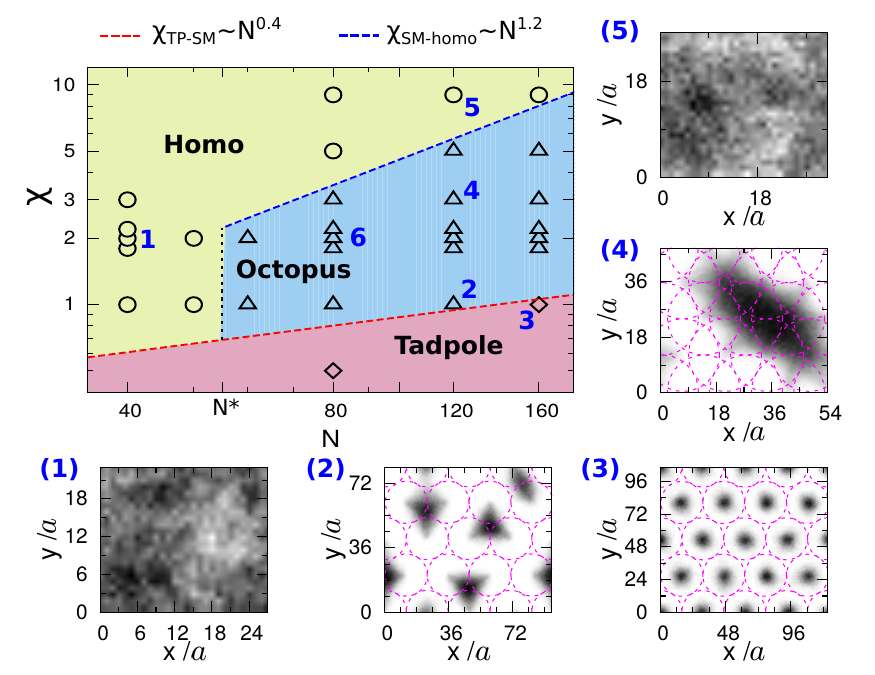}%\vspace*{-5.5pt}
\caption{
Structural diagram of polyelectrolyte brush simulated at $q_{3+}=1.0$ 
as a function of chain length $N$ and overlap parameter $\chi$. 
Typical lateral density profiles of brushes are shown together. 
Dashed lines are phase boundaries $\chi \sim N^{0.4}$ and $\chi \sim N^{1.2}$, 
which are derived from a scaling argument. 
The vertical phase boundary formed due to the competition between chain fluctuations and trivalent ion-mediated interaction is marked at $N^*\approx 50-60$ between homogeneous and octopus phases. 
The density profile at state point (6) is shown in Fig.~\ref{betaHRAT}b-(3).
}
\label{pDiag}
\end{figure*}

{\bf Structural diagram. }
At $q_{3+}=1.0$, the stoichiometric condition at which the most heterogeneous brush structures emerge, 
the effects of chain size ($N$) and %grafting density ($\sigma$ or $\chi$) 
chain overlap ($\chi$), which is related to the grafting density $\sigma$ as $\chi \sim \sigma R_{g,xy}^{2}$,
on the morphology of brush are further studied (Group C in Table \ref{sysinfo}). %\ref{sysinfo}). 
As depicted in Fig.~\ref{pDiag} (see also Fig.~\ref{cfgs}), three distinct phases are present.  
As a trivial case, tadpole-like condensates are expected if $\sigma$ is very low 
(see the brush of $N=160, \chi=1.0$, Fig.\ref{pDiag}-(3)). 
When the polyelectrolyte chains are either short ($N=40$, Fig.\ref{pDiag}-(1)) 
or grafted with a very high density $\chi > 5.0$ (Fig.\ref{pDiag}-(5)), 
the morphology of brush condensates is homogeneous. 
With $1 \leq \chi \leq 3$, octopus-like surface micelle domains are formed  
at an intermediate range of $\chi$ (Fig.\ref{pDiag}-(2) or (4)). 
Assuming that the collapsed phase of polyelectrolyte brush with trivalent ions is reasonably mapped onto the collapsed polymer brush in a poor solvent, 
we treated the counterion-mediated interactions as an effective attraction and used the scaling argument \cite{2005Pereira214908} to describe the phase boundaries in Fig.\ref{pDiag}.

{\it Boundary between tadpole and surface micelle}: 
The free energy of individual tadpole composed of a single chain, 
like the case of $N=160$ and $\chi=1.0$ (Fig.\ref{pDiag}-(3)), 
is dominated by its surface energy, 
$F_{\text{TP}} \sim \gamma  R^{2}$, 
where $\gamma$ is an effective surface tension mediated by the salt concentration, 
and $R$ is the size of a tadpole-like structure. 
Assuming that the surface energy is around $k_{B} T$ per monomer, $\gamma \sim \xi k_{B}T/{a^{2}}$, where the parameter $\xi$, which is maximized at $q_{3+}=1$, denotes a scale of counterion mediated-effective attraction between chain monomers,    
and $R \sim a N^{\nu}$ for a collapsed globule of polyelectrolyte chain, 
$F_{\text{TP}}$ is then given by 
\begin{equation}
\frac{F_{\text{TP}}}{k_{B}T} \sim \xi N^{2\nu}.
\end{equation}

The free energy of a bigger fused micelle composed of three neighboring chains, 
like the case of $N=120$ and $\chi=1.0$ (Fig.\ref{pDiag}-(2)), 
can be expressed as a summation of a surface energy term $\gamma ((3N)^{\nu} a)^{2}$, 
and a elastic energy term $3 k_{B}T d^{2}/(N_{s} a^2)$ 
as a result of stretching $N_{s} = d/a$ monomers in each chain to reach the surface micelle. 
Taken together, it yields 
\begin{equation}
\frac{F_{\text{FM}}}{k_{B}T} \sim \xi(3N)^{2\nu} + 3 \frac{d}{a}.   
\end{equation}
Finally, balance between $F_{\text{FM}}$ and $F_{\text{TP}}$ leads to a critical grafting distance $d_c$ (or grafting density, $\sigma_c$), 
$\sigma_{c} \sim d_{c}^{-2} \sim \xi^{-2}N^{-4\nu}$. 
Therefore, the phase boundary between tadpole and fused surface micelle has the following scaling behavior in terms of the chain overlap parameter ($\chi \sim \sigma R_{g,xy}^2\sim \sigma N^{1.7}$) and $N$. 
\begin{align}
\chi\sim \xi^{-2}N^{1.7-4\nu}.
\label{eqn:scaling1} 
\end{align}

{\it Boundary between surface micelle and homogeneously collapsed layer}: 
In a similar way, the free energy per unit area of a homogeneously collapsed layer, 
like the case of $N=120$ and $\chi=9.0$ (Fig.\ref{pDiag}-(5)), 
is $f_{\text{homo}} = 2 \gamma$,  
\begin{equation}
\frac{f_{\text{homo}}}{k_{B}T} \sim \xi\frac{2}{a^2}.
\end{equation}
where the factor 2 arises because there are two interfaces with the collapsed polymer; one at the grafting points and another at the  interface with the bulk.  

Meanwhile, for an octopus-like domain of size $R_{n}$, 
containing $n$ chains within a surface area defined by $\sim R^2_{c}$,  
the surface energy is 
$F_{n,\text{surf}}\sim \gamma R_{n}^{2} \sim \xi k_{B} T (n N)^{2\nu}$, 
and the elastic energy is 
$F_{n,\text{elastic}}\sim n k_{B}T R_{c}^{2}/(N_{s} a^2) = n k_{B}T R_{c}/a = k_{B}T \sigma R_{c}^{3}/a$ 
where we used the relationships of $R_c=N_s\times a$ and $\sigma=n/R_c^2$. 
Hence the total free energy, per unit area, in octopus-like condensate with $n$ arms, is 
\begin{align}
\frac{f_{\text{octo}}}{k_{B}T} &= \frac{1}{k_BT}\frac{F_{n,\text{surf}}+F_{n,\text{elastic}}}{R_c^2}\nonumber\\
&=\frac{ \xi (n N)^{2\nu} + \sigma R_{c}^{3}/a }{R_{c}^{2}} = \frac{\xi (\sigma N)^{2\nu}}{R_{c}^{2-4\nu}} + \frac{\sigma R_{c}}{a}. 
\end{align}
Minimization of the free energy, $f_{\text{octo}}$, with respect to $R_{c}$ 
gives a minimum value of $f_{\text{octo}}(R_c^*) \sim N^{\frac{2\nu}{3-4\nu}} \sigma^{\frac{2-2\nu}{3-4\nu}}$, 
at the following size $R_c^*$: 
\begin{align}
R^*_{c} \sim N^{\frac{2\nu}{3-4\nu}} \sigma^{\frac{2\nu-1}{3-4\nu}}.
\end{align} 
By balancing $f_{\text{octo}}$ with $f_{\text{homo}}$, we also obtain the critical grafting density between these two phases, and hence the chain overlap parameter $\chi$ which scales with $N$ and $\xi$ as follows. 
\begin{equation}
\chi \sim \xi^{\frac{1-2\nu}{1-\nu}}N^{1.7+\frac{\nu}{\nu-1}}.
\label{eqn:scaling2} 
\end{equation}

To recapitulate, for $\nu=1/3$, which is the scaling exponent for collapsed polymer, 
the phase boundary between tadpole and fused micelle (surface micelle) is  
$\chi_{\text{TP-SM}}\sim \xi^{-2} N^{0.4}$, 
and the phase boundary between homogeneously collapsed brush and octopus-like surface micelle is 
$\chi_{\text{SM-homo}}\sim \xi^{1/2} N^{1.2}$. 
Therefore, the surface micelle phase is enclosed by these two boundaries 
and its area in Fig.\ref{pDiag} is maximized when $q_{3+}=1$, where the strength of effective attraction between monomers reaches its maximum value ($\xi\le \xi_{\text{max}}$). 
The vertical boundary (the dotted line in Fig.\ref{pDiag}) between homogeneous and octopus phases at $N^*\approx 50-60$ is due to large fluctuations of short chains which predominate over the counterion-induced attraction between neighboring chains and prevent them from aggregating into a metastable octopus-like surface micelle. 
\\

%Fig.8
\begin{figure*}[t]
\centering\includegraphics[width=1.7\columnwidth]{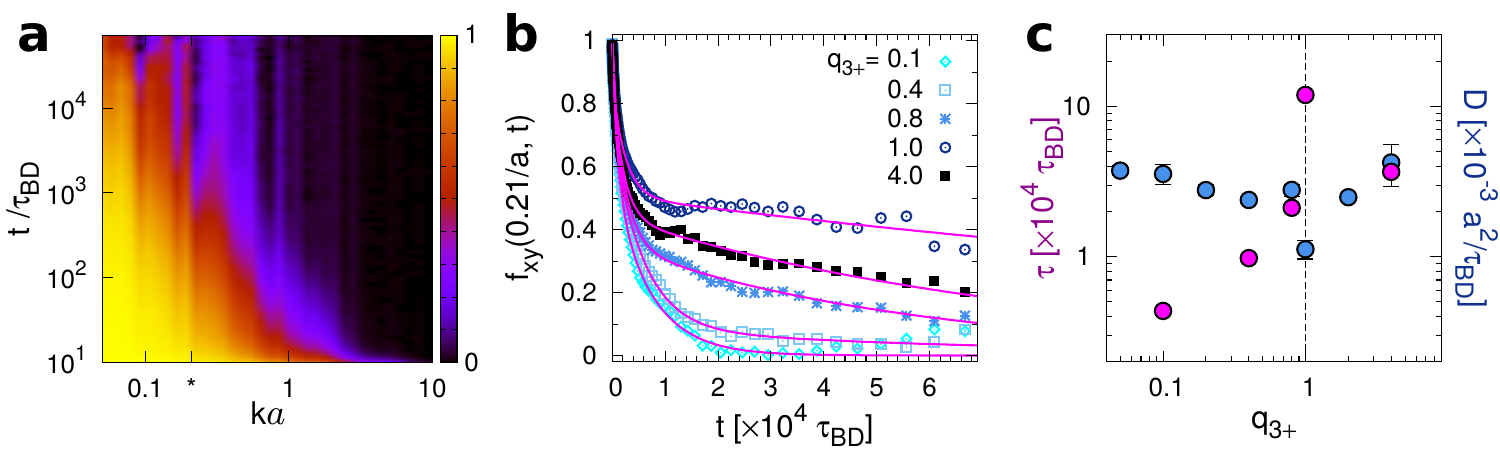}%\vspace*{-5.5pt}
\caption{Dynamics of brush condensates. 
(a) The heatmap of normalized intermediate scattering function $f_{xy}(k,t)$ 
for a polyelectrolyte brush with $N=80$, $\chi=2.0$, and at $q_{3+}=1.0$. 
(b) $f_{xy}(k=0.21a^{-1},t)$ %$f_{xy}$ at $k=0.21$ versus simulation time 
for bushes at $q_{3+}=0.1,0.4,0.8,1.0,4.0$. 
Solid lines are fittings of $f_{xy}(k,t)$ to 
a multi-exponential function. 
(c) The relaxation time $\tau$ and diffusion coefficient of trapped trivalent ions as a function of $q_{3+}$. 
%(c) Diffusion coefficient of trapped trivalent ions as a function of $q_{3+}$. 
}
\label{isf}
\end{figure*} 

{\bf Dynamics of brush condensate. }
The intermediate scattering function for chain monomers \cite{2006Hansen}, 
\begin{align}
F_{xy}(k,t) = \Big\langle \Big \langle \frac{1}{N_{m}^2} \sum_{m=1}^{N_{m}} e^{i \vec{k} \cdot \vec{r}_{m}(t+t_{0})} \sum_{n=1}^{N_{m}} e^{-i \vec{k} \cdot \vec{r}_{n}(t_{0})} \Big \rangle_{|\vec{k}|} \Big \rangle_{t_{0}}
\label{eq-isf}
\end{align}
is evaluated using the trajectories generated from Brownian dynamics simulations. 
$F_{xy}(k,t)$ describes the conformational relaxation of brush chains on different spatial scales. 
The heatmap of $f_{xy}(k,t) \left[= F_{xy}(k,t)/F_{xy}(k,0)\right]$, 
for the brush with $N=80, \chi=2.0$ is plotted in Fig.~\ref{isf}a. 

The fast decaying behavior of $f_{xy}(k,t)$, i.e., on small length scales ($ka>1$),  
indicates that the monomers inside the collapsed surface micelles are fluidic, 
and that the trivalent ion-mediated polymer-polymer attractions are still transient. 
In contrast, the relaxation dynamics of polymer brush in a larger length scale ($ka < 1$) is rather slow. 
The scattering function $f_{xy}(k=0.21a^{-1}, t)$, %$f_{xy}(t)$ calculated at $k=0.21$, 
which probes the dynamics of brush on the length scale of surface micelle size ($R=2\pi/k\approx 30 a$), 
exhibits slow decaying behavior, 
which is fitted to a multi-exponential function (Fig.~\ref{isf}b) 
or to a stretch-exponential function $\sim \exp(-(t/\tau)^\beta)$ 
with $\beta < 1$ (Fig.~S1). 
Together with the exponent of stretch-exponential $\beta$ less than 1, which is a quintessential marker of glassy dynamics \cite{Kang2015PRL},  
the non-monotonic dependence of $\tau$ on $q$ should be noted (Fig.~\ref{isf}c). 
The relaxation dynamics of octopus-like surface micelle at $q_{3+} \approx 1$ 
is $\sim 30$ fold slower than that at $q_{3+}=0.1$. 
Given that the size of polyelectrolyte chain we simulated is finite ($N=80$), 
it is expected that the relaxation time of octopus-like surface micelle would effectively be divergent at a mesoscopic chain size $N\gg1$ as in Ref.  \cite{2014BarZiv4945}. 
Once trapped in a dynamically metastable state, the heterogeneous configurations of surface micelle are maintained for an extended amount of time. 

At $q_{3+}=1.0$, when the relaxation dynamics of polyelectrolyte chain is the slowest, the trivalent ions trapped in the brush region exhibit the slowest dynamics as well; the diffusion coefficient of trivalent ions trapped in the brush region, which can be calculated from an ensemble averaged mean square displacement, is minimal ($D_{\text{ion}}^{\text{brush}}\sim 10^{-3}$ $a^2/\tau_{\text{BD}}$, Fig.~\ref{isf}c) and is only $\sim 1/100$ of that in the bulk ($D_{\text{ion}}^{\text{bulk}}\sim 10^{-1}$ $a^2/\tau_{\text{BD}}$).
When $q_{3+} > 1.0$, the effective correlation between trapped ions and monomers becomes smaller than the value at $q_{3+}=1.0$ (see Fig.\ref{ssf}c), giving rise to a greater diffusivity (Fig.\ref{isf}c).

\section*{DISCUSSION}
At high trivalent counterion concentration ($q_{3+} \gtrsim 1.0$), dense polyelectrolyte brushes composed of long chains ($\chi > 1.0$ and $N > 40$) collapse to spatially inhomogeneous brush condensates.  
The octopus-like surface micelle domains, demarcated by chain-depleted grain boundaries, become most evident at the ionic condition of $q_{3+}=1$.  
Our structural diagram (Fig.\ref{pDiag}), the accompanying scaling arguments, and dynamics of brush condensates assessed by intermediate scattering functions (Eq.\ref{eq-isf}, Fig.\ref{isf}) qualitatively capture the essence of the experimental observations by Bracha \emph{et al.} \cite{2014BarZiv4945}.   
In their experiment, DNA brushes underwent a transition into macroscopic dendritic domains, 
only if the spermidine concentration $C_\text{Spd}$ 
was greater than the critical value $C_\text{Spd}^{*}$ ($\approx 61~\mu\text{M}$). 
Just like the vertical phase boundary at $N^*\approx 50-60$ demonstrated in the structural diagram of our study, 
the minimal chain length of $\sim 1$ kb was required for the growth of dendritic domains. 
It was also observed that for $C_\text{Spd} > C_\text{Spd}^{*}$, the collapsed domains were smaller in a denser brush of shorter chains, 
which is consistent with the prediction from our scaling argument $R_{c} \sim N^{0.4} \sigma^{-0.2}$. The scaling relationship of $R_{c} \sim N^{0.4} \sigma^{-0.2}$ further explains the various size of octopus-like micelle domain depicted in (2), (4), (6) in Fig.~\ref{pDiag}.
According to Fig.~\ref{pDiag}, 
the heterogeneous morphology of collapsed brush is more likely to be acquired for large $N$; however, when the grafting density is extremely high ($\chi\gg 1$) crossing the phase boundary of $\chi_{\text{SM-homo}}\sim N^{1.2}$, the morphology of brush condensate becomes homogeneous regardless of $N$.   

It is noteworthy that there have been a large amount of studies on a collection of free polyelectrolytes in solution \cite{Stevens1999PRL,sayar2006EPL,Fazli2007PRE}. 
Yet, the geometric constraint of the brush system with one end being grafted makes substantial difference. While the two systems share similarity in that the attraction between two polymer chains is initiated and mediated by counterions, which in a certain condition for free polyelectrolytes leads to a macrophase segregation through infinite bundling \cite{Stevens1999PRL,sayar2006EPL,Fazli2007PRE}, the very constraint on the one termini in brush system limits the size of aggregates leading to lateral microphase segregation, the domain size of which is determined by $N$ and $\sigma$ (or $\chi$), i.e., $R_{c} \sim N^{0.4} \sigma^{-0.2}$. 

A few cautionary remarks are in place in comparing our studies with the existing dsDNA biochip experiments.
Compared with the DNA brush experiment \cite{2014BarZiv4945}, 
the simulated brush is more than an order of magnitude shorter, 
but with a higher grafting density. 
Importantly, we have not taken into account the finite chain stiffness of double-stranded DNA \cite{2001Stevens130} by reasoning that the ratio of contour ($L$) and persistence lengths ($l_p$) of dsDNA used in biochip experiment ($L/l_p\sim \mathcal{O}(10^2)$) is large enough to ignore the effect of chain stiffness on  overall conclusion of our study. However, in reality the local chain stiffness does affect the local co-alignment of brush condensate, which is propagated and amplified into a macroscopic scale.
As a consequence, the chain organization inside octopus-like domain differ from that inside the dendritic domain in Ref. \cite{2014BarZiv4945}, where tens to hundreds chains are co-aligned in a width of 20 -- 100 nm and tightly packed in the hexagonal columnar order \cite{2001Kenneth14925,2014BarZiv4945}. 
In addition, different models for ions and polyamine compounds, such as their size and charge distribution \cite{1997Nordenskiold4335,Yu2016Macromolecules} and spatial inhomogeneity in grafting density could modulate the critical salt concentration for maximal heterogeneity.  

Furthermore, although our study was carried out through a specific protocol of preparation, that is by adding trivalent salt to the chain-free zone (bulk) and waiting for the system to be equilibrated, it may well be that the result of this exercise changes with the protocol of how the trivalent ions are added. As reported by the experiments \cite{2014BarZiv4945,2016BarZiv142}, different modes of incorporating spermidine into brush (e.g., the rate by which trivalent counterions are added) lead to slightly different critical concentrations, different collapse dynamics, and different morphologies. 
All these kinetic effects of spermidine addition are closely linked to the ultra-slow dynamics of brush condensates at $q_{3+}\approx 1$, underscored in Fig.\ref{isf}, and are of great interest for the future study. 

To recapitulate, we study indicates that when a condition is met, lateral heterogeneity develops in pericellular matrices \cite{2009TanakaS671} as well as in DNA biochip. 
The results of our study also imply that the local compactness of DNA in interphase chromosome is manipulated  
through dynamic controls over the DNA density and the biogenic polyamine concentration \cite{yoo2016NatComm}, 
which in turn regulates DNA transcription activities. 
\\

\section*{ACKNOWLEDGMENTS}
We are grateful to Bae-Yeun Ha and Wonmuk Hwang for useful comments. We thank the Center for Advanced Computation in KIAS for providing supercomputing resources.  

%\bibliography{psRNA,mybib1,additionalRefs}
%\bibliographystyle{achemso}
\bibliography{rev_condmat}
%\end{thebibliography}

\clearpage 

\setcounter{figure}{0}  
\setcounter{equation}{0}
\makeatletter 
\renewcommand{\thefigure}{S\@arabic\c@figure}
\renewcommand{\theequation}{S\@arabic\c@equation}
\makeatother 

\section{Supporting Information}
%\\

\begin{figure}[h!]
\centering\includegraphics[width=.96\columnwidth]{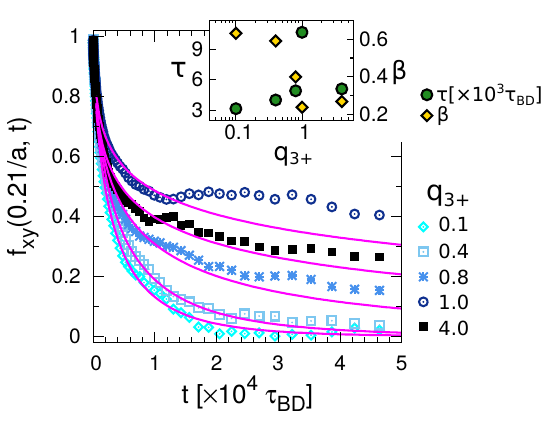}%\vspace*{-5.5pt}
\caption{
Lateral intermediate scattering function $f_{xy}(0.21a^{-1},t)$ %at $ka=0.21$ versus simulation time 
for bushes with various salt concentrations $q_{3+}=0.1,0.4,0.8,1.0,4.0$. 
Solid lines are fittings of $f_{xy}(k,t)$ to 
a stretched exponential function $\sim \exp(-(t/\tau)^\beta)$, 
where $\tau$ and $\beta$ are plotted as a function of $q_{3+}$ in the insert. 
}
\label{isf-SI}
\end{figure} 

\end{document}